\documentclass{IEEEcsmag}

\usepackage[colorlinks,urlcolor=blue,linkcolor=blue,citecolor=blue]{hyperref}

\usepackage{upmath}

\usepackage{psfrag}
\usepackage{subfigure}
\usepackage{url}
\usepackage{stfloats}
\usepackage{cite}
\usepackage{amsmath}
\usepackage{amssymb}
\usepackage{amsfonts}
\usepackage{graphicx}
\usepackage{fancybox}
\usepackage{color}
\usepackage{algorithm}
\usepackage{algorithmic}
\usepackage{multirow}
\usepackage{upgreek}
\usepackage{booktabs}
\usepackage{threeparttable}
\usepackage{latexsym}
\usepackage{bm,array}
\usepackage{graphicx}
\usepackage{float}
\usepackage{mathtools}
\usepackage{commath}
\usepackage{amsmath,lipsum}
\usepackage{cuted}%%\stripsep-3pt

\jvol{XX}
\jnum{XX}
\paper{8}
\jname{Computer}
\pubyear{2022}

\begin{document}

%\sptitle{Department: Head}
%\editor{Editor: Name, xxxx@email}

\title{Space Broadband Access: The Race Has Just Begun}

\author{Yiming Huo}
\affil{\emph{Senior Member, IEEE} \\University of Victoria, BC, Canada\\Correspondence: yhuo@ieee.org \\ \\This article has been accepted by IEEE, with DOI:10.1109/MC.2022.3160472}

%\author{T. C. Author, III}
%\affil{Third Affiliation}

%\markboth{Department Head}{Paper title}

\begin{abstract}
Recent years have witnessed an exponential growth of the commercial space industry, including rocket launch, satellite network deployment, private space travel, and even extraterrestrial colonization. Several trends are predicted in this unprecedented transition to an era of space-enabled broadband access.\\


\end{abstract}

\maketitle

\chapterinitial{I. The Introduction} Connecting the unconnected and delivering high-quality internet access are the continuous efforts made by internet service providers (ISPs) \cite{Khaturia:Connect}. In spite of the fact that the low-latency and high-speed internet service is usually easier to access in modernized urban areas through diverse ways and with multiple service providers, facilitating high-quality internet service in rural and remote areas is more challenging \cite{Koziol:Money}. This is mainly because the physical barriers of the natural environments can lead to the significantly increased cost of deployment and maintenance of network infrastructure. Moreover, fiber optic repeaters are needed to amplify the signal that usually gets largely attenuated after traveling over tens of kilometers, which makes the total installations cost up to 14,000 US dollars per kilometer \cite{Koziol:Money}.

On the other hand, the recent boom in the commercial space industry \cite{Victor:Private} has gradually demonstrated its impact on altering the game of internet service providing. Several private space corporations have deployed or planned to deploy a large number of low earth orbit (LEO) satellites for broadband access. It is predicted that 2022 and beyond will be the time of ``technological singularity'' when significantly more LEO satellites are to be launched, deployed, and planned to broaden and enhance the space-enabled broadband access to more customers and regions. Meanwhile, more participants joining this market will escalate the competition and technological evolution, eventually improving the overall quality of service (QoS) and user experience (UX).  
%\section{COPYRIGHT AND CLEARANCE}
%All CS Magazine authors must obtain clearance from IEEE Computer Society before submitting the final manuscript. The ``\href{https://apps.na.collabserv.com/wikis/home?lang=en-us#!/wiki/W18e544042a85_4b63_915a_1d1ed2cf8338/page/Publication\%20clearance}{Publication Clearance}'' wiki provides details about the procedure. Computer Society employees must use the \href{https://mc.manuscriptcentral.com/cs-ieee}{ScholarOne  Manuscripts Clearance} System to obtain publication approval.
\section{II. Prediction 1 - multiple boosters will accelerate the scale-up of space broadband networks}

Since the launch of the satellite Telstar built by Bell Labs in 1962, commercial communication satellites have been growing fast, particularly during the 1990s when several companies enabled the satellite-delivered internet. However, there was some temporary regression due to the immature business model, technical challenges, relatively high cost, etc. Compared to more than two decades ago, today has seen many revolutionary and critical enablers of spaceborne broadband access as one essential part of 6G networks \cite{Giordani:6G}.

\begin{figure*}
\centering
\centerline{\includegraphics[width=30pc]{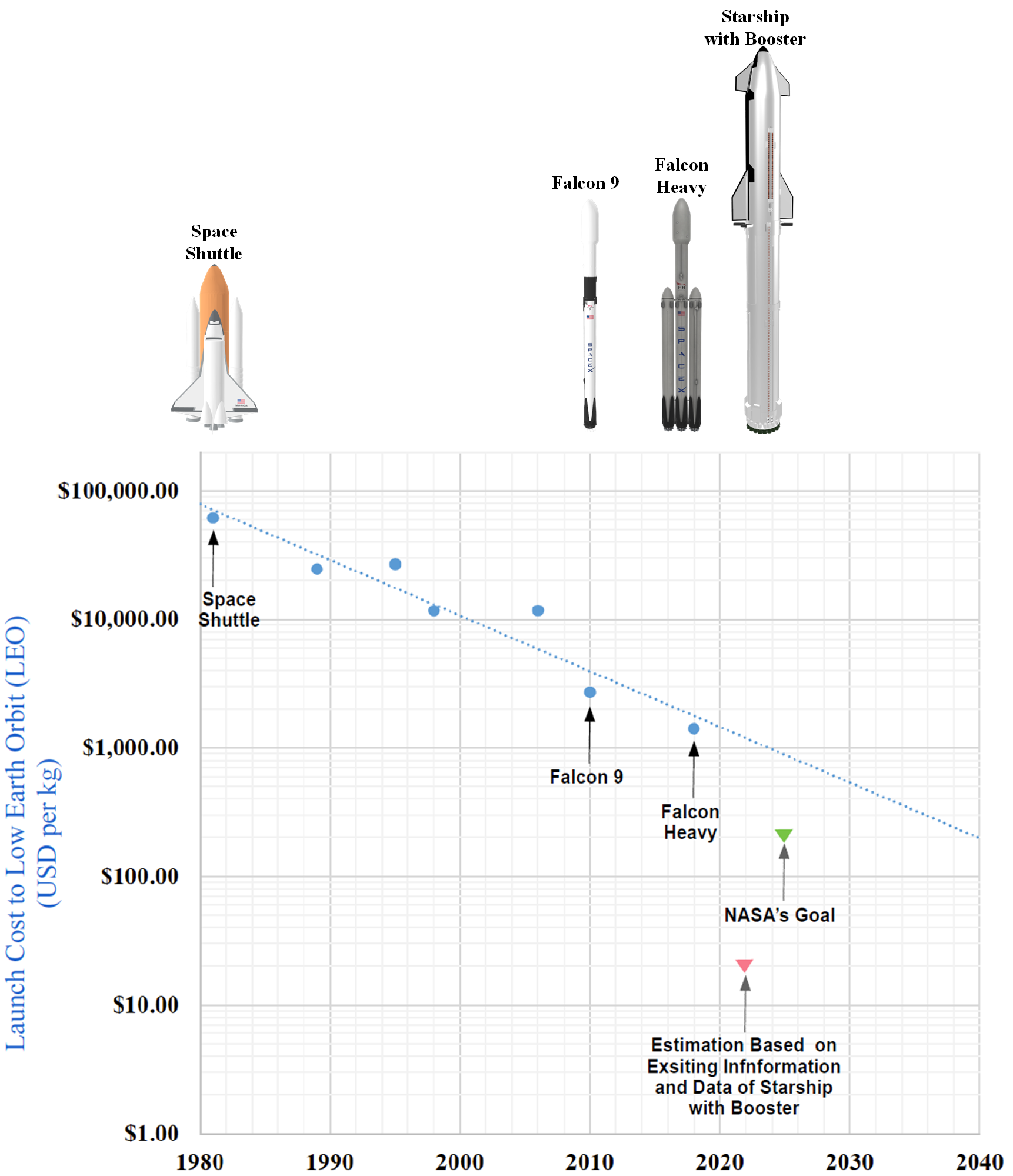}}
\caption{Low earth orbit launch cost change over time (Rockets and Space Shuttle photo credit: AllThingsSpace, via https://sketchfab.com/sunnychen753 CC BY-SA 4.0).}\label{fig:launch}
\end{figure*}

\begin{figure*}
\centering
\centerline{\includegraphics[width=30pc]{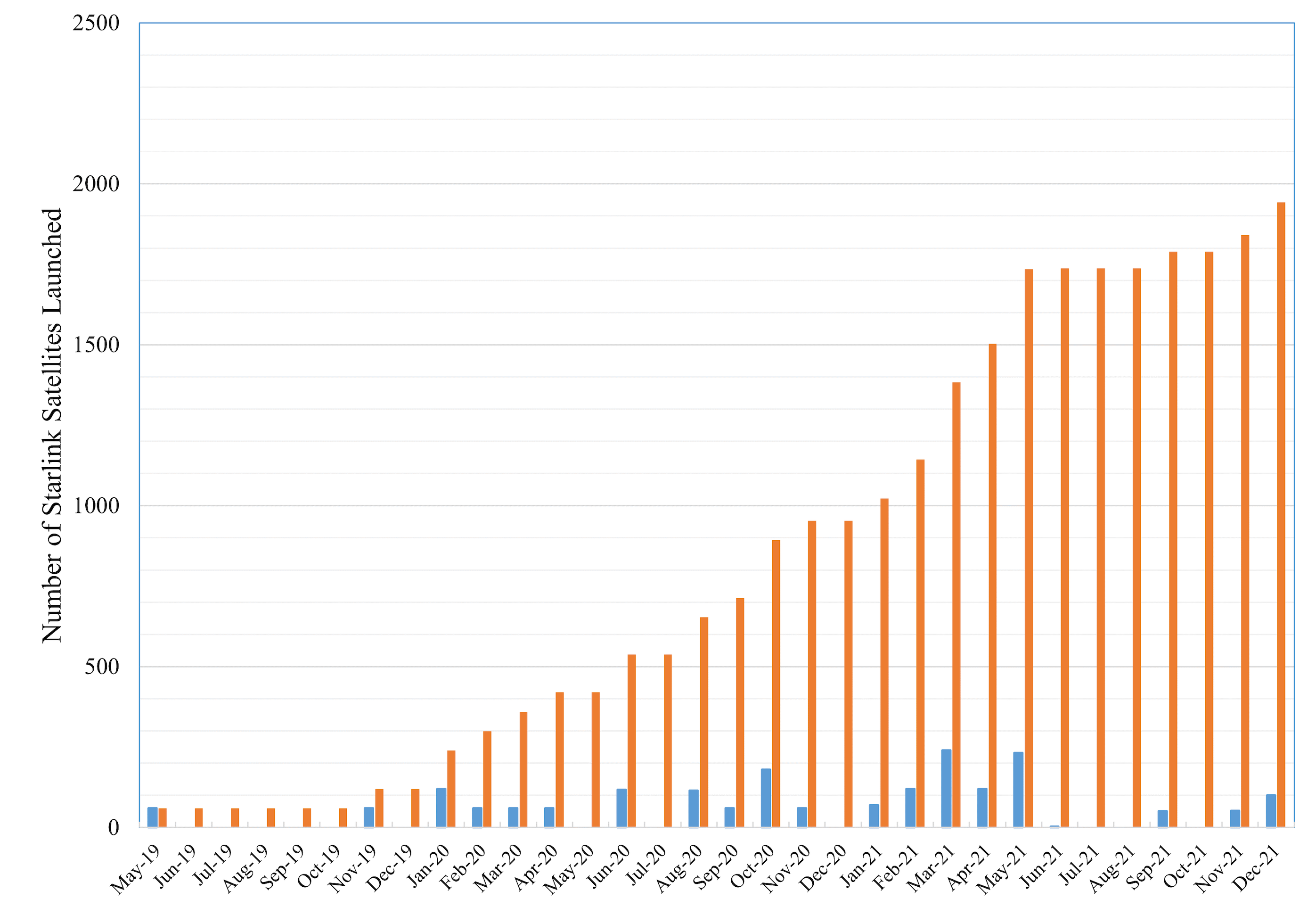}}
\caption{Starlink satellites launch information on a monthly basis.}\label{fig:LaunchNo.}
\end{figure*}

\subsection{Rocket Launch}
The space industry has been conventionally a capital-intensive risky business \cite{Musk:Business}, but things have constantly changed due to disruptive factors such as private capital investment in recent years. Since its first commercial mission in 2013, SpaceX has been leading the commercial launch service. One critical milestone was marked in 2015 when SpaceX successfully launched and relanded its Falcon 9 rocket \cite{Evan:SpaceX}, which paves the road of the launch vehicles' reusability and even lower launch costs. 

As illustrated in Fig.~\ref{fig:launch}, the launch cost data of the last several decades \cite{Jones:Cost} are reported and labeled. The blue round dots indicate the inflation-weighted costs of several representative LEO launches dating back to 1981, when the space shuttle was launched for the first time. The latest verifiable launch cost data is Falcon Heavy in 2018, 1,420 USD per kilogram for the payload. In particular, Jan. 24, 2021, witnessed a record 143 satellites launched on one Falcon 9 rocket. A trendline in blue dots is plotted based on the de facto launches. As observed, the trendline is a bit further away from the green triangular indicating NASA's goal set to make the launch cost reduced to 100 USD/pound (220 USD/kg) in 2025 \cite{NASA}. Nevertheless, the scheduled first orbital flight of SpaceX's Starship that can deliver more than 100 tons of payload to the LEO \cite{Starship} might bend the trend line significantly downwards in 2022. According to Mr. Elon Musk, the launch cost is estimated to be 2 million USD per launch \cite{Mann:SpaceX}, which can significantly reduce the cost to 20 USD/kg.  

%Moreover, another California-headquartered startup, Astra Space, successfully launched its 13-meter LV0007 rocket and delivered a demonstration payload to the orbit nearly 500 km. The company designed its launch vehicle to be highly responsive and flexible to make Astra's entire launch system transportable and mobile. As predicted, the launch cost will continue to drop in 2022.

\subsection{Satellite}
As of December 2021, there are a total of 1,783 Starlink satellites currently consisting of a non-geostationary (NGSO) constellation to serve customers globally, at altitudes ranging from 440--570 km. Since the first mission of deploying Starlink satellites in May 2019, there have been 33 launch missions carrying a total of 1,942 Starlink satellites (until Dec. 31, 2021), and many other companies' satellites in rideshares to the LEO. The monthly number of launched Starlink satellites is illustrated in Fig.~\ref{fig:LaunchNo.}. The year 2021 has recorded a total number of 1,822 Starlink satellites launched.    

Except for the first launch mission of 60 Starlink satellites V0.9, which has a mass of 227 kg, all the remaining Starlink satellites have a mass of 260 kg with Ku and Ka-band antennas. In addition, Hall-effect thrusters using low-cost krypton gas as the reaction mass are equipped on all satellites for orbit raising and station-keeping. One solar array is used to power supply the entire satellite, including transceivers working at Ku and Ka bands and inter-satellite laser links on some satellites. It is worth mentioning many Starlink Gateway stations have been built globally to connect satellites that are not equipped with laser links to terrestrial optic infrastructure. Hundreds of more such stations are planned \cite{Arevalo:SpaceX} although the future inter-satellite link (ISL) enabled by laser can remove some need for ground stations.    

From the communication system perspective, phased-array technology has been widely used to design the satellite, gateway station, and user terminal \cite{SpaceX:Gen2}. The currently operating Starlink first-generation system, or Gen1, can enable a theoretical throughput of 17--23 Gbps (20 Gbps on average) per satellite, with latency from 25 to 35 ms. As of December 2021, the theoretical total throughput of 1,783 Starlink satellites could achieve roughly 35.66 Tbps, enabling 356,600 users to have a downlink speed of 100 Mbps simultaneously, or 1,783,000 users to watch YouTube 4K videos (20 Mbps) at the same time. Eventually, 4,425 Gen1 NGSO satellites will be deployed to provide an estimated total throughput of 88.5 Tbps. With a launch pace recorded in 2021, the Gen1 constellation is predicted to be finished before the end of 2023.

Furthermore, in the proposed Starlink Gen2 system, a new frequency and channelization plan submitted to the Federal Communications Commission (FCC) is presented and compared with Gen1 in Table~\ref{Table:Frequency}. As observed, the Gen2 system will add more Ku and Ka bands frequencies to user downlink/uplink. Moreover, using the V-band frequencies with 5 GHz contiguous bandwidth at both downlink and uplink is also proposed \cite{SpaceX:Gen2}. Furthermore, in the Gen2 system, SpaceX has offered to deploy 29,996 Starlink LEO and very low earth orbit (VLEO) satellites at altitudes from 328--614 km \cite{SpaceX:Gen2A}.   

\begin{table}
\newcommand{\tabincell}[2]{\begin{tabular}{@{}#1@{}}#2\end{tabular}}
 \centering
\footnotesize
\caption{Frequency and Channelization Plan of Starlink System.}\label{Table:Frequency}
%\vspace{-0.5cm}
\begin{threeparttable}
\begin{tabular}{|c|c|c|} \hline
\tabincell{c}{\textbf{Type of Link} \\ \textbf{and Transmission}\\ \textbf{Direction}} & \tabincell{c}{\textbf{Gen1} \\ \textbf{Frequency} \\ \textbf{Ranges (GHz)}} & \tabincell{c}{\textbf{Gen2} \\ \textbf{Frequency} \\ \textbf{Ranges (GHz)}} \\
\hline
\tabincell{c}{User Downlink \\ Satellite-User \\ Terminal} & 10.7--12.7 & \tabincell{c}{10.7--12.75$^{1}$ \\ 17.8--18.6 \\ 18.8--19.3 \\ 19.7--20.2} \\
\hline
\tabincell{c}{Gateway Downlink \\ Satellite to \\ Gateway} & \tabincell{c}{17.8--18.6 \\ 18.8--19.3} & \tabincell{c}{17.8--18.6 \\ 18.8--19.3 \\ 71.0--76.0} \\
\hline
\tabincell{c}{User Uplink \\ User Terminal\\ to Satellite} & 14.0--14.5 & \tabincell{c}{12.75--13.25$^{2}$ \\ 14.0--14.5 \\ 28.35--29.10 \\ 29.5--30.0} \\
\hline
\tabincell{c}{Gateway Uplink \\ Gateway to Satellite} & \tabincell{c}{27.5--29.1 \\ 29.5--30.0} & \tabincell{c}{27.5--29.1 \\ 29.5--30.0 \\ 81.0--86.0} \\
\hline
\tabincell{c}{TT\&C Downlink} & \tabincell{c}{12.15--12.25 \\ 18.55--18.60} & \tabincell{c}{12.15--12.25 \\ 18.55--18.60} \\
\hline
\tabincell{c}{TT\&C Uplink} & {13.85--14.00} & {13.85--14.00} \\
\hline
\end{tabular}
%\vspace{-1.0cm}
    \begin{tablenotes}
        \footnotesize
        \item[1] SpaceX does seek authority to provide service in the United States using the 12.7--12.75 GHz band, but proposes to use that spectrum in other areas of the world where allowed \cite{SpaceX:Gen2}. 
        \item[2] At this time (when this application is filed), SpaceX seeks authority to use this band in the United States only with individually-licensed earth stations. No such limitations would apply outside the U.S. \cite{SpaceX:Gen2}.
      \end{tablenotes}
\end{threeparttable}
\end{table}

The constant improvement of the antenna and integrated circuit (IC) designs (e.g., \cite{Rastic:Satcom}) and semiconductor fabrication processes has enabled a more powerful, compact, and energy-efficient satellite system \cite{Burleigh:Survey}. Therefore, the Gen2 system can enable much higher throughput than the Gen1 due to the deployment of more satellites to facilitate more spectrum resources and higher performance per satellite and ground station. It is estimated the total throughput of the Gen2 constellation can be ten times as the Gen1, which can bring the total throughput of Gen1 plus Gen2 more than 1000 Tbps, or 1 Pbps.   

To summarize the prediction for the Gen2 constellation: first, its deployment may commence after the complete deployment of the Gen1, which may finish before the end of 2023; second, if SpaceX could increase the average monthly number of launched LEO satellites from the current level (152 on average in 2021) to 300 or 600, it would take 100 or 50 months to reach full deployment. Then, we may see the complete deployment of both Gen 1 and Gen 2 systems earliest around 2028. 

\subsection{Artificial Intelligence and Cloud/Edge Computing}
In addition, some new technologies will serve as essential boosters of space broadband deployment. Compared to two decades ago, artificial intelligence (AI) \cite{Fourati:AI} and cloud/edge computing today play a crucial role in enabling high-performance satellite broadband connectivity and services such as beam-hopping, interference managing, satellite traffic control \cite{Kato:SAGINs}, image processing, mapping \cite{Mahdianpari:Mapping}, and computation offloading \cite{Tang:Offloading}. 

In May 2021, Google won a deal to supply networking resources to Starlink satellites whose internet will rely on Google's private fiber-optic network to make quick connections to Google's cloud services \cite{Novet:Google}. Instead of outsourcing computing power or data storage to data centers, SpaceX will install ground stations at Google data centers to facilitate fast, secure, and intelligent connectivity. 

Earlier in 2020, Microsoft announced working with SpaceX to deliver satellite connectivity between field-deployed assets and cloud resources globally via Starlink. They planned to further connect Starlink with Microsoft's global network, including Azure edge devices \cite{Jennifer:Space}. 

Similarly, Amazon Web Services (AWS), which provides advanced on-demand cloud computing platforms and APIs (application programming interfaces), can be vertically integrated with Amazon's own LEO satellite constellation Project Kuiper \cite{Koziol:Amazon}. Therefore, it is predicted that more cloud computing and AI-empowered satellite constellation will be seen/announced in 2022 and later years.

\section{III. Prediction 2 - all humanity will have to address the challenges of enabling space broadband}
 
\begin{figure} 
\begin{minipage}{.5\linewidth}
\centering
\subfigure[]{\includegraphics[scale=.5]{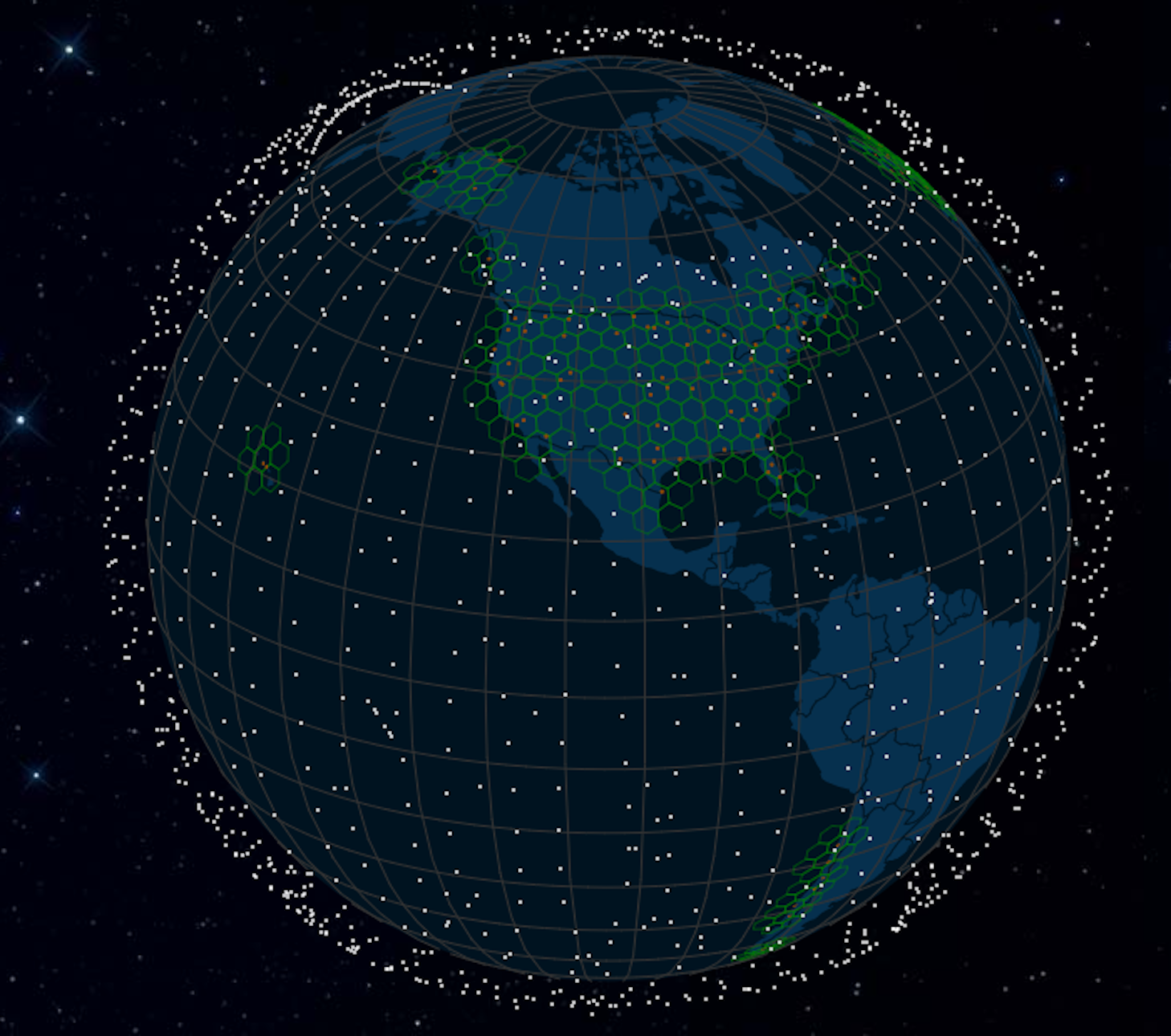}}
\end{minipage}
\begin{minipage}{.5\linewidth}
\centering
\subfigure[]{\includegraphics[scale=.49]{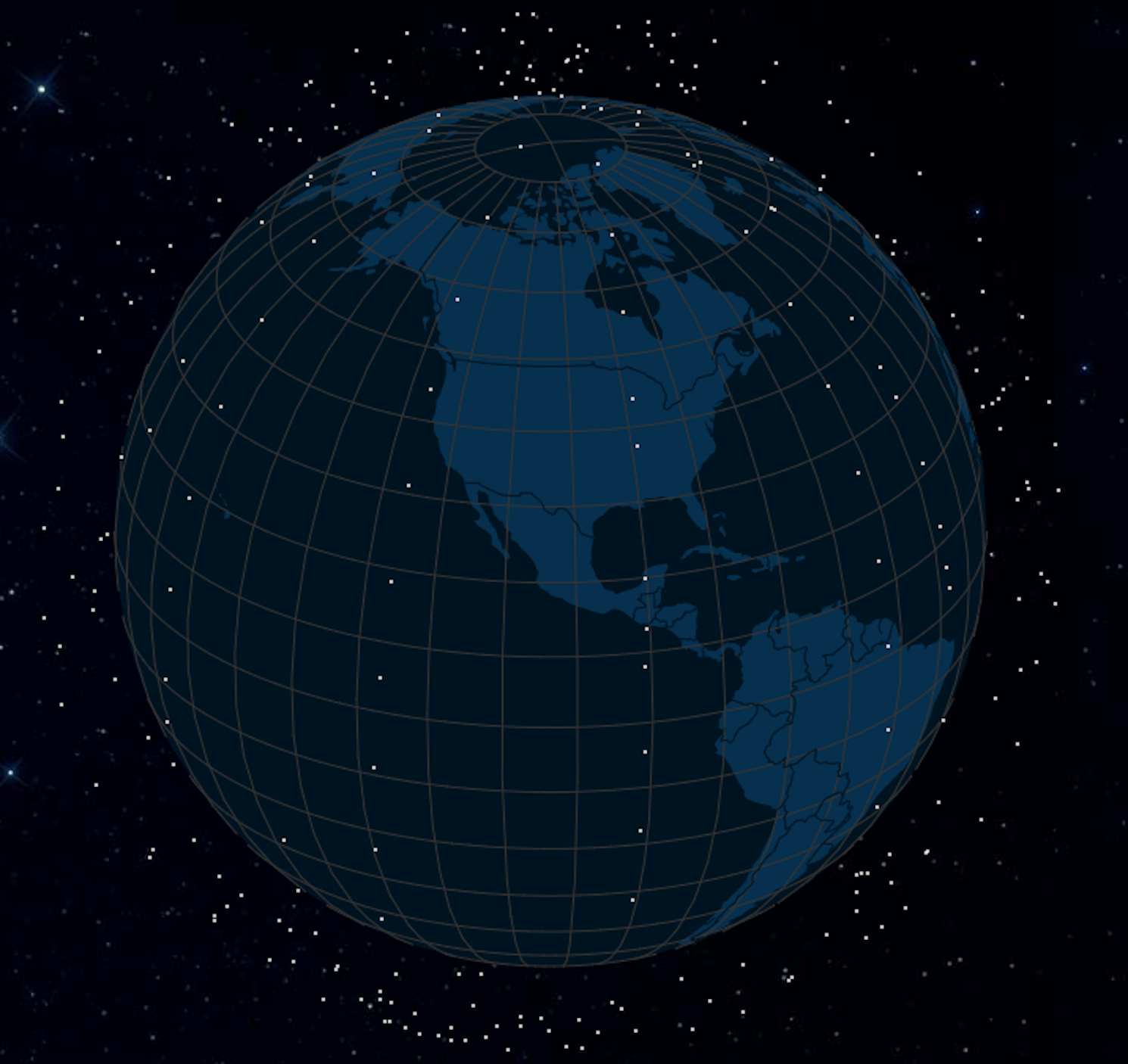}}
\end{minipage}\par\medskip
\caption{Illustration of real-time orbital and location information of Satellites in (a) SpaceX Starlink constellation and (b) OneWeb constellation, using \cite{Tracker}.} \label{fig:Orbit}
\end{figure}
 
\subsection{Competition}
Although SpaceX now owns the largest LEO satellite constellation, several other players are catching up in this area. In terms of the launched NGSO satellites quantity, the London-headquartered communications company, OneWeb (Network Access Associates Ltd) comes in second place by recording a total of 394 LEO satellites deployed in the orbital plane of 1,200 km, which is about half of its initial constellation accommodating 648 satellites. OneWeb's intended markets are primarily to businesses, governments, and clusters of communities, which is different than Starlink focusing on individual domestic customers. The comparison of the two satellite constellations is illustrated in Fig.~\ref{fig:Orbit}, where Starlink's mega-constellation is deployed at lower orbits with higher density. In comparison, OneWeb needs to place satellites to higher orbits to maintain good coverage. 

Moreover, the Ottawa-based satellite communications company, Telesat, has deployed 298 LEO satellites for its Telesat Lightspeed$^{\text{TM}}$ network aiming at connectivity solutions for corporations and governments such as internet/wireless backhaul and inflight/maritime connectivity. Furthermore, several other companies such as Amazon, Boeing, are seeking to build their own broadband satellite constellations, as detailed in Table~\ref{Table:Constellation}.

\begin{table}
\newcommand{\tabincell}[2]{\begin{tabular}{@{}#1@{}}#2\end{tabular}}
 \centering
\footnotesize
\caption{Comparison of Broadband LEO Satellite Constellations (as of Dec. 2021).}\label{Table:Constellation}
%\vspace{-0.5cm}
\begin{threeparttable}
\begin{tabular}{|c|c|c|} \hline
\tabincell{c}{\textbf{Company and} \\ \textbf{Constellation} \\ \textbf{Name}} & \tabincell{c}{\textbf{Number of} \\ \textbf{Satellites} \\ \textbf{Deployed}} & \tabincell{c}{\textbf{Number of} \\ \textbf{Satellites} \\ \textbf{Proposed}} \\
\hline
\tabincell{c}{SpaceX \\ Starlink} & 1,783 & 41,939 \\
\hline
\tabincell{c}{OneWeb (NAA Ltd) \\ OneWeb}  & 394 & 6,372 \cite{CNBC} \\
\hline
\tabincell{c}{Telestat \\ Telestat Lightspeed}  & 298 & 1,373 \cite{CNBC} \\
\hline
\tabincell{c}{Amazon \\ Project Kuiper}  & 0 & 7,774 \cite{CNBC} \\
\hline
\tabincell{c}{Boeing}  & 0 & 5,789 \cite{CNBC} \\
\hline
\end{tabular}
%\vspace{-1.0cm}
%    \begin{tablenotes}
%        \footnotesize
%        \item[1] SpaceX does seek authority to provide service in the United States using the 12.7--12.75 GHz band, but proposes to use that spectrum in other areas of the world where allowed \cite{SpaceX:Gen2}. 
%        \item[2] At this time (when this application is filed), SpaceX seeks authority to use this band in the United States only with individually-licensed earth stations. No such limitations would apply outside the U.S. \cite{SpaceX:Gen2}.
%      \end{tablenotes}
\end{threeparttable}
\end{table}

When looking at the bigger picture, more players and competition can accelerate the technological advances and the related supply chains' upgrading, which results in cost decline. The research of ARK invest has unveiled Wright’s law applies to satellite bandwidth \cite{ARK}. Since 2004 the cost of satellite bandwidth has dropped 7,500-fold from \$300,000,000/Gbps to \$40,000/Gbps, and it could fall another 40-fold during the next five years to ~\$1,000/Gbps. In 2022 and later years, it is predicted that solid growth in the subscription of satellite-enabled broadband access will be observed.

\subsection{Coexistence and Integration}
Along with a foreseeable fast-booming space-enabled broadband market comes the challenges. An interference resulting in performance degradation among different LEO satellite constellations is a significant one. Although the available contiguous spectrum is comparatively more abundant at Ku/Ka/V-band than the sub-6 GHz region, all the mentioned companies work to maximize the throughput of each constellation. More and more satellites operating at these bands may cause the increase of both inter-constellation interference and intra-constellation interference. For example, the interference could appear between Starlink's and OneWeb's downlink in the Ku-band. Moreover, the user terminal could interfere with other satellites when connecting with one satellite in the uplink, even within the same constellation\cite{Jia:Uplink}. 

In addition, integrating satellite-enabled broadband networks with conventional terrestrial networks \cite{Burleigh:Survey} and other non-terrestrial networks (consisting of aerial vehicles and high-altitude platforms (HAPs)) to form space-air-ground integrated networks (SAGINs) poses many challenges. They include interference, mobility and constellation management \cite{Giordani:6G}, task scheduling, resource allocation, etc. 

Consequently, a basket of effective interference-avoidance/mitigation strategies, such as interference coordination, resource allocation management, beam scheduling and management, dynamic bandwidth channelization, etc., are predicted to be developed, verified, and deployed to cope with the challenges mentioned above. Moreover, more cost-effective AI techniques will shed light on achieving a higher performance gain such as in \cite{Kato:SAGINs}. 

\subsection{Space Traffic and Junk}
More concerns of space collision and space junk have been raised recently, when the space in the Earth's orbits has seen drastically increased congestion. In May 2021, engineers in NASA spotted a 5 mm wide hole in one robotic arm of the International Space Station (ISS), which is created by an unknown piece of space junk \cite{Nature:Collision}. Sometime earlier in July 2018, the European Space Agency (ESA) had to fire the thrusters on the USD 162-million CryoSat-2 spacecraft to boost it into a higher orbit to avoid a possible collision with a piece of space debris \cite{Witze:Quest}. These close calls may indicate a higher probability and frequency of similar incidents to happen in 2022 and beyond. Furthermore, they highlight the urgency of developing a sustainable global framework for managing space traffic by governments and companies. Such a framework may include space collision prevention and safe de-orbiting/disposal strategy, requiring technical and legislative solutions. It is predicted that such a framework built on global joint efforts could take effect within several years. 

\subsection{Beyond Earth}
The space-enabled wireless communication and internet services can be extended to another celestial body. In October 2020, NASA selected NOKIA to build the first-ever cellular network on the Moon which is orbiting Earth at an average distance of 384,400 km and takes 1.2822 seconds for the wireless signal to finish a one-way trip. With its pioneering innovations, Nokia Bell Labs will construct and deploy the first ultra-compact, low-power, space-hardened, end-to-end LTE communications networks on the lunar surface in late 2022 \cite{Nokia}. This unprecedented project will lay the crucial communications foundation for NASA's Artemis program intended to return astronauts to the Moon by 2024 \cite{Witze:Moon}. This project marks another milestone for space-enabled internet services. It needs to overcome the challenges such as the long communication distance, high latency, and more severe interference from cosmic rays and solar flare. 

Furthermore, based on Elon Musk's vision that humans may become a multi-planetary species \cite{Elon:Mars}, a more daring but still realistic prediction is that space-enabled broadband access will be deployed on other planets. Mars can be the first after humanity sets up a sustainable presence on it \cite{Elon:Mars}. Considering that the astronomical positions of Mars and Earth vary over time, more challenges have to be overcome. Fortunately, various nations have accumulated abundant valuable experiences from tens of Mars missions. As soon as humanity establishes a settlement on Mars (possibly some time the end of this decade or 2030s), space-enabled broadband access will be available between Mars and Earth.  

\section{IV. CONCLUSION}
In 2022 and later years, humanity's digital life will experience more technological transformation, such as broadband access facilitated or enhanced by next-generation satellite communications and mega-constellations. Unprecedented opportunities are unfolded into the community, thanks to the pioneering scientists, engineers, and entrepreneurs. The continued launch cost reduction, spectrum/energy efficiency increase, advances of AI and cloud/edge computing, supply chain upgrading will serve as major key catalysts to further lower the overall infrastructure cost, accelerate the deployment, and eventually cover and benefit more global users. More vertical applications on top of satellite constellations will become possible. On the other hand, the unavoidable challenges are predicted to emerge when more participants compete, and different networks need to be integrated, such as radio interference management, space traffic control, sustainable environmental compatibility, etc. Moreover, space-enabled broadband access will be extended to the Moon, Mars, and other celestial bodies in the future. To make this transition more smooth, efficient, and beneficiary, the involved entities and individuals, may seek technical solutions and collaboration and establish the related legislation.   

\section{ACKNOWLEDGMENT}

Author thanks Dr. Chris Ng, the IEEE Future Networks Massive MIMO Working Group Co-Chair and Director of Blue Danube Systems, and Dr. Ashutosh Dutta, the Founding Co-Chair of IEEE Future Networks Initiative, for the inspiration and support.

\begin{IEEEbiography}{Yiming Huo} is a research associate with the Department of Electrical and Computer Engineering, University of Victoria, Canada. He received Ph.D. in electrical engineering from the same university. Dr. Huo received multiple paper and society awards and the honor of IEEE \emph{Access} Most Popular Article in 2017. He worked in the industry, including Ericsson, ST-Microelectronics, and Apple Inc. 

He is a Senior Member of the IEEE, an Associate Editor for the IEEE \emph{Access}, and Guest Editor of \emph{Signals}. Dr. Huo served as Organizing Committees (TPC Co-Chair, Publicity Co-Chair) of the IEEE Future Networks First Massive MIMO Workshop in 2021 and the Publication Chair of IEEE PACRIM in 2019. His current research interests include next-generation wireless systems, space technology, space exploration, astronomy, internet of things, and machine learning. %Contact him at yhuo@ieee.org.

\end{IEEEbiography}

\end{document}